\documentclass[useAMS,usenatbib]{mn2e}
\usepackage[dvips]{graphicx}
\usepackage{amsmath}
\usepackage{amsfonts}
\usepackage{amssymb}

\title[Transit timing effects due to an exomoon]{Transit timing effects due to an exomoon}
\author[David M. Kipping]{David M. Kipping$^{1}$\thanks{E-mail:
d.kipping@ucl.ac.uk}\footnotemark[1]\\
$^{1}$Department of Physics and Astronomy, University College London, \\
       Gower Street, London WC1E 6BT, UK}
\begin{document}

\date{Accepted 2008 September 20. Received 2008 August 18; in original form 2008 July 15}

\volume{392} \pagerange{181--189} \pubyear{2009}

\maketitle

\label{firstpage}

\begin{abstract}
As the number of known exoplanets continues to grow, the question as to whether such bodies harbour satellite systems has become one of increasing interest.  In this paper, we explore the transit timing effects that should be detectable due to an \emph{exomoon} and predict a new observable.  We first consider transit time variation (TTV), where we update the model to include the effects of orbital eccentricity.  We draw two key conclusions: \\
1) In order to maintain Hill stability, the orbital frequency of the exomoon will always be higher than the sampling frequency. Therefore, the period of the exomoon cannot be reliably determined from TTV, only a set of harmonic frequencies.\\
2) The TTV amplitude is $\propto M_S a_S$ where $M_S$ is the exomoon mass and $a_S$ is the semi-major axis of the moon's orbit. Therefore, $M_S$ \& $a_S$ cannot be separately determined.

We go on to predict a new observable due to exomoons - transit duration variation (TDV).  We derive the TDV amplitude and conclude that its amplitude is not only detectable, but the TDV signal will provide two robust advantages:\\
1) The TDV amplitude is $\propto M_S a_S^{-1/2}$ and therefore the ratio of TDV to TTV allows for $M_S$ and $a_S$ to be separately determined.\\ 2) TDV has a $\pi/2$ phase difference to the TTV signal, making it an excellent complementary technique.

\end{abstract}


\section{Introduction}

Over 300 exoplanets have been discovered to date with detection rates escalating (see http://exoplanet.eu by J. Schneider).  The detections have been so far biased towards large bodies and the smallest transiting planet to date is still Neptune-sized, for Gliese 436b (\citet{gil07}).  Current instruments cannot yet detect transiting Earths but transit time variation (TTV) could offer a way to bring sensitivity down to sub Earth-mass level (\citet{hol05} and \citet{ago05}).  Given the large number of moons in our own Solar System, it is reasonable to postulate that satellites are common around exoplanets.

Another profound motivation for looking for exomoons is that they are likely to be terrestrial in nature, based upon our own Solar System, and hence one would propose that exomoons could be more habitable environments than the host of extrasolar giant planets (EGPs) so far discovered.

Photometric detection of a moon is likely to be exceptionally challenging.  An exomoon is likely to be sub-Earth sized, based upon our Solar System, and so even a specialized transit space-based telescope like COROT will struggle to spot the signature (\citet{sar99}).  An additional problem lies in the fact that much of the time the exomoon will not appear to be its orbital distance from the planet, but some fraction of it, depending on the orbital phase of the exomoon during transit.  The moon can effectively hide behind the planet or in front of it.  This makes disentangling the photometric signature exceptionally difficult and was outlined by \citet{sar99} and \citet{cab07}.

Previously, several different authors (\citet{sza06}, \citet{sar99} and \citet{sim07}) have noted that transit timing variations (TTV) could be used to indirectly detect the presence of such exomoons.  In the first half of this paper, we update the model for the TTV effect by including orbital eccentricity.  We compare our formulation to the previously proposed mathematical treatment and demonstrate our model reduces to the original analytic equations for zero eccentricity.

However, the crucial problem with TTV is that the amplitude of the signal is proportional to both the exomoon mass, $M_S$, and orbital separation, $a_S$.  \citet{for07} referred to this obstacle as the `inverse problem', in regard to transit timing effects.  In \S3.1 we show that the exomoon's period cannot be reliably determined from TTV and hence $a_S$ remains an unknown.  Ergo, one cannot establish the mass of the exomoon without assuming a value for the orbital separation.  It is therefore clear that a strong desideratum is a secondary method which can complement TTV and remove this degeneracy and this constitutes the focus of the second part of our paper.

By considering the transit duration, we predict a new observable timing effect due to an exomoon, which we label as transit duration variation (TDV).  We find that the amplitude of this timing signal is of the same order of magnitude to the TTV signal and indeed often larger.  The effect is also predicted to be $\propto M_S a_S^{-1/2}$.  Hence, the ratio of TDV to TTV allows for the mass of the exomoon to be found without assuming an orbital distance.  In addition, TDV is $\pi/2$ out-of-phase with TTV, making it an ideal complementary method for exomoon detection.

\section{TTV Amplitude due to an Exomoon}
\subsection{Outline of the model}

In the first half of this paper, we aim to update the model for the transit timing variation (TTV) signal due a transiting exoplanet with a single satellite of mass $M_S$ and non-zero orbital eccentricity.  In this work, we consider the variation of the mid-transit point of the \emph{planetary} transit, $T_{MID}$.  This is in contrast to \citet{sim07} who consider the planet and moon combined transit.  Throughout our discussion, we also make the assumption that planet-moon orbital plane is co-aligned with the planet-star orbital plane at $i=90^{\circ}$.

In our case, the planet orbits the barycentre of the planet-moon system with a semi-major axis of $a_W$ where $W$ denotes wobble.  The fact that $a_W > 0$ means that the time between the planet being at the mid-transit point and the barycentre being at the mid-transit point is, in general, non-zero, and this is the origin of the TTV effect.

Consider the projected distance between the planet and the planet-moon barycentre to be given by $x_2'$, as illustrated in figure 1.  The TTV effect will be given by $x_2'$ divided by the $\hat{x}_2'$-direction component of the barycentre's orbital velocity around the star, given by $v_{B\bot}$.  Since $x_2'$ is a function of the planet's true anomaly around the planet-moon barycentre, $f_W$, so too is the TTV effect:

\begin{equation}
\textrm{TTV}(f_W) = \frac{x_2'(f_W)}{v_{B\bot}}
\end{equation}

\begin{figure}
\begin{center}
\includegraphics[width=8.4 cm]{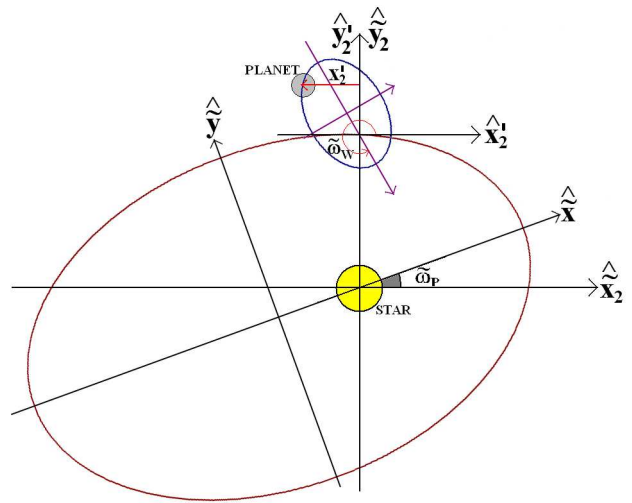}
\caption{\emph{Cartoon of the star-planet-moon system.  In this schematic, the observer lies at $\tilde{y_2} = +\infty$ and the exomoon is not shown, but the wobble of the planet due its presence is represented by the smaller ellipse.  The origin of the $x_2'$-$y_2'$ scale denotes the position of the planet-moon barycentre, which itself orbits the star in the $\tilde{x_2}$-$\tilde{y_2}$ scale.}} \label{fig:fig1}
\end{center}
\end{figure}

In the case of a circular orbit, we expect the TTV signal to have a sinusoidal nature, and infact the peak-to-peak amplitude of the wave may be unambiguously defined as $2 a_W/v_{B\bot}$.  In the case of eccentric orbits, we expect non-sinusoidal waveforms, due to Kepler's Equation.  Peak-to-peak amplitudes can become ambiguous in such cases, and so we choose to use the root-mean-square (rms) amplitude definition, which is valid for all waveforms.  For a simple circular orbit, the rms amplitude of the TTV signal, $\delta_{TTV}$, will be given by:

\begin{equation}
\delta_{TTV}(\textrm{\emph{circular}}) = \frac{a_W}{\sqrt{2} \cdot v_{B\bot}}
\end{equation}

Where the distance $a_W$ is small enough that we may assume $v_{B\bot}$ is a constant over the time-scale of the TTV effect.

Let us now extend the analysis to include the effect of exomoon orbital eccentricity, $e_S$ and position of pericentre, $\varpi_S$, as well as planetary eccentricity $e_P$ and position of pericentre, $\varpi_P$.  In appendix A, we derive the general equation for the rms amplitude of the TTV effect due to an exomoon to be given by:

\begin{equation}
\delta_{TTV} = \frac{1}{\sqrt{2}} \cdot \frac{a_P^{1/2} a_S M_S M_{PRV}^{-1}}{\sqrt{G (M_* + M_{PRV})}} \cdot \frac{\zeta_{T}(e_S,\varpi_S)}{\Upsilon(e_P,\varpi_P)}
\end{equation}

where
\begin{align}
\zeta_{T} &= \frac{(1-e_S^2)^{1/4}}{e_S} \sqrt{e_S^2 + \cos(2 \varpi_S) (2 (1-e_S^2)^{3/2} - 2 + 3 e_S^2)} \\
\Upsilon &= \cos\Big[\arctan\Big(\frac{-e_P \cos\varpi_P}{1+e_P \sin\varpi_P}\Big)\Big] \cdot \sqrt{\frac{2 (1+e_P \sin\varpi_P)}{(1-e_P^2)}-1}
\end{align}

Where $a_P$ is the semi-major axis of the exoplanet's orbit around the central star, $a_S$ is the semi-major axis of the exomoon's orbit around the exoplanet, $G$ is the gravitational constant, $M_*$ is the mass of the host star and $M_{PRV} = M_S + M_P$ i.e. the combined mass of the exomoon and exoplanet respectively\footnote{$M_{PRV}$ is the mass of the planet as measured by radial velocity surveys, and thus includes the mass of satellites}.

In the case of $e_S \rightarrow 0$ we have $\zeta_T \rightarrow 1$ and similarly for $e_P \rightarrow 0$ we have $\Upsilon \rightarrow 1$.  Thus we may compare our equation to that derived by previous authors, who assumed circular orbits, and expect to produce the same result.  \citet{sar99} predicted the peak-to-peak amplitude of an exomoon around an exoplanet with orbital period $P_P$ to be given by equation (6).

\begin{equation}
\Delta t_{sar99} \sim 2 a_S M_S M_P^{-1} \times P_P (2 \pi a_P)^{-1}
\end{equation}

Using Kepler's Third Law, it is trivial to show that this is equivalent to our expression, for circular orbits and $M_P \gg M_S$, except for a factor of $2\sqrt{2}$, which comes from the fact we use an rms definition of amplitude, rather than a peak-to-peak definition.  Thus we have confirmed that our general expression for the TTV effect due to an exomoon is equivalent to the circular equations first derived by \citet{sar99}.  For completion, we note that the TTV amplitude may be written purely in terms of the masses of the three bodies and the period of the exoplanet (see appendix A, equation (A28) \& (A29)).

The effect of eccentricity is discussed in more detail in \S3.3 and typical waveforms produced are shown later in figure 3, \S4.2.  In \S4.2, Table 1 gives a list of the expected TTV rms amplitudes due to a $1 M_{\bigoplus}$ exomoon around some of the most promising exoplanet candidates.

\subsection{Permitted range for $a_S$}

The orbital radius of any satellite around a planet must lie somewhere between the Hill radius, $d_{max}$, and the Roche limit, $d_{min}$, to maintain stability\footnote{Note, we use the rigid body Roche limit for simplicity}.  We express $a_S$ by assuming that it is equal to some fraction, $\chi$, of the Hill radius, $d_{max}$.

\begin{align}
a_S &= \chi \cdot d_{max} \\
d_{max} &= a_P \cdot \Big( \frac{M_P}{3 M_*} \Big)^{1/3} = a_P \cdot \Big( \frac{M_{PRV}-M_S}{3 M_*} \Big)^{1/3} \\
d_{min} &= R_P \cdot \Big(2 \frac{\rho_P}{\rho_S}\Big)^{1/3}
\end{align}

Where $a_P$ is the semi-major axis of the planet around the host star, $R_P$ is the planetary radius, $\chi$ is some real number between 0 and 1, and $\rho$ denotes density.

$\chi$ may be further constrained by noting that \citet{bar02} estimated $\chi \lesssim 1/3$ and \citet{dom06} estimated $\chi \lesssim 1/2$.  This is because the Hill sphere is just an approximation, and in reality other effects, like radiation pressure or the Yarkovsky effect, can perturb a body outside of the sphere.  We choose to use the conservative choice of $\chi \lesssim 1/3$ and combining this limit with the Roche limit, and rewriting in terms of planetary period, $P_P$ for $M_* \gg M_P$, we can estimate:

\begin{equation}
\Big(\frac{18 \pi}{G P_P^2 \rho_S}\Big)^{1/3} \lesssim \chi \lesssim \frac{1}{3}
\end{equation}

\citet{val06} offer a way of rewriting $\rho_S$ in terms of just the mass of the exomoon by adopting terrestial models of the internal structure for such bodies.  If we use $R_S \propto M_S^{0.27}$, then we can estimate $\rho_S \sim 0.1 M_S^{0.19}$.

\begin{equation}
\frac{1}{186} \Big(\frac{M_S}{M_{\oplus}}\Big)^{-0.063} \Big(\frac{P_P}{\textrm{1 day}}\Big)^{-2/3} \lesssim \chi \lesssim \frac{1}{3}
\end{equation}

\section{Implications}
\subsection{The high frequency nature of an exomoon's TTV}

If we make the approximation that $M_* \gg M_{PRV}$ and $M_{PRV} \gg M_S$, and employ Kepler's Third Law, we may provide an estimate for the ratio of the exomoon to planet orbital period, which can be shown (see appendix B) to be:

\begin{equation}
\frac{P_S}{P_P} \simeq \sqrt{\frac{\chi^3}{3}}
\end{equation}

Since $\chi \leq 1$ we will always be in the regime where $P_S < P_P$.  Taking $\chi \sim 1/3$ gives a rough estimate of $P_S/P_P \sim 1/9$ and so the frequency of the signal is $\nu_S \sim 9 P_P^{-1}$.  However, the Nyquist frequency will be given by one half of the sampling rate ($0.5 P_P^{-1}$), which represents the maximum frequency we can resolve without aliasing.

The usual technique for detecting signals within data is to employ a periodogram, but this method will suffer from aliasing in the exomoon case.  Therefore, we can only derive a set of harmonic frequencies which the exomoon's orbit could exhibit.  We therefore conclude $P_S$ cannot be reliably determined from the TTV signal.

\subsection{The limitation of TTV in determining $M_S$}

As originally pointed by \citet{sar99}, the major problem with TTV is that one cannot determine the mass of the exomoon without making an assumption on the distance at which the moon orbits the planet.  Using equation (3), it is possible to write the TTV amplitude as a function of purely the exomoon properties:

\begin{equation}
\delta_{TTV} \propto M_S a_S
\end{equation}

Therefore we can effectively only determine the moment of the exomoon.  If one knows the period of the exomoon, then it is trivial to derive $a_S$ using Kepler's Third Law, but as seen in \S3.1, $P_P$ cannot be reliably ascertained.

This crucial limitation of exomoon TTV makes mass estimation unfeasible and the best we can ever do is merely provide evidence for an exomoon within a large mass range.  This severe limitation will be resolved later in this paper.

\subsection{The effect of eccentricity}

The effects of orbital eccentricity on the TTV amplitude are all absorbed into the two parameters, $\zeta_T(e_S,\varpi_S)$ and $\Upsilon(e_P,\varpi_P)$.  If we increase $e_S$ from zero to unity, regardless of what value $\varpi_S$ takes, $\zeta_T$ decreases below 1 and hence the TTV amplitude will always decrease.  This implies exomoon detections are biased towards satellites on circular orbits.

The situation is more complicated for $e_P$, where a non-zero $e_P$ makes the TTV amplitude significantly increased or decreased depending on $\varpi_P$.  In figure 2, we plot $\Upsilon^{-1}$ as a function of $\varpi_P$ for several different eccentricities and find that for $e_P>0$, the most favourable position of pericentre is $\varpi_P \sim 3 \pi/2$.  However, we point out a recent study by \citet{kan08} which predicts such exoplanets to possess a low transit probability.

For the eccentric transiting exoplanets GJ436b, XO-3b, HAT-P-2b and HD17156b, $\Upsilon^{-1}$ takes values 1.01, 1.03, 0.94 and 0.47 respectively.  However, the stability of satellites around eccentric exoplanets remains unclear.

\begin{figure}
\begin{center}
\includegraphics[width=8.4 cm]{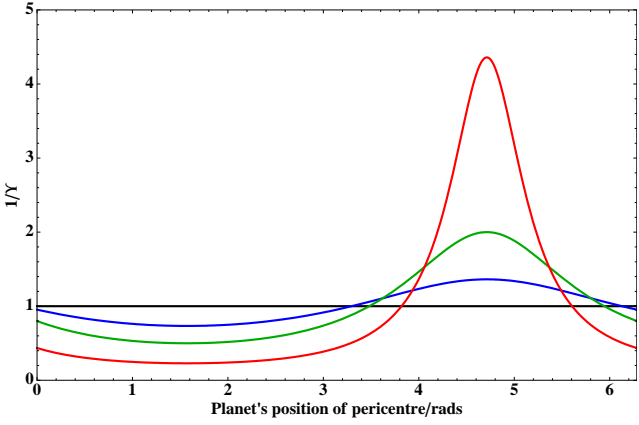}
\caption{\emph{Behaviour of $\Upsilon^{-1}$ as a function of $e_P$ and $\varpi_P$, which is effectively the factor by which the TTV is enhanced by for $e_P >0$.  Positions of pericentre near $270^{\circ}$ produce consistently enhanced TTV amplitudes.  The flat line corresponds to $e_P = 0$ and each progressively larger amplitude wave is for $e_P = 0.3$, $0.6$ and $0.9$ respectively.}} \label{fig:fig2}
\end{center}
\end{figure}

\section{Transit Duration Variation (TDV)}
\subsection{The TDV due to an exomoon}

Transit duration variation (TDV) is the periodic change in the duration of a transit ($\tau$) over many measurements, where we define $\tau$ as the time between the 1$^{\textrm{st}}$ and 4$^{\textrm{th}}$ contact points.  It has previously been discussed as a possible test of general relativity by \citet{pal08}.  In this discussion, we consider the TDV due to an exomoon and conclude that it should produce a detectable signal.

We first consider that the duration of a transit is inversely proportional to the projected velocity of the planet across the star, $v_{P\bot}$, and make the following assumptions:

\begin{itemize}
\item[{\tiny$\blacksquare$}] The projected velocity of the planet-moon barycentre during the transit, relative to the star, does not vary significantly over the time-scale of the transit duration.
\item[{\tiny$\blacksquare$}] The projected velocity of the planet during the transit, relative to planet-moon barycentre, does not vary significantly over the time-scale of the transit duration.
\item[{\tiny$\blacksquare$}] The orbital inclinations of the planet ($i = 90^{\circ}$) and exomoon do not vary from orbit to orbit.
\item[{\tiny$\blacksquare$}] We do not consider additional perturbing bodies in the system.
\end{itemize}

\begin{equation}
\tau \propto 1/v_{P\bot}
\end{equation}

In this case, any transit duration variation must be solely due to the variation of the velocity.  For a single companion exomoon, the velocity of the planet will have two components:

\begin{equation}
v_{P\bot} = v_{B\bot} + v_{W\bot}
\end{equation}
where $v_{B}$ is the velocity of the planet-moon barycentre around the host star and $v_{W}$ is the velocity of the planet around the planet-moon barycentre, i.e. the \emph{wobble} of the planet due its companion satellite.

It is clear that $v_{W\bot}$ will be significantly different for each transit unless $P_P/P_S$ is some low-order integer.  Sometimes $v_{W\bot}$ will be additive to the barycentre's velocity and sometimes subtractive resulting in shorter and longer transit durations respectively; an effect we label TDV.

The factor by which $\tau$ will vary must be equal to the ratio of the velocities $R = v_{W\bot}/v_{B\bot}$ and from this starting point the TDV amplitude may be shown (see appendix C2) to be equal to:

\begin{equation}
\delta_{TDV} = \sqrt{\frac{a_P}{a_S}} \cdot \sqrt{\frac{M_S^2}{M_{PRV} (M_{PRV} + M_*)}} \cdot \frac{\bar{\tau}}{\sqrt{2}} \cdot \frac{\zeta_D(e_S,\varpi_S)}{\Upsilon(e_P,\varpi_P)} 
\end{equation}

where
\begin{equation}
\zeta_D(e_S,\varpi_S) = \sqrt{\frac{1 + e_S^2 - e_S^2 \cos(2 \varpi_S)}{1-e_S^2}}
\end{equation}

In an analogous way to the TTV amplitude, this may be re-written in terms of simply the masses in the system, and this equation may be found in the appendix C2, equations (C28) and (C29).  We also point out that this effect has the following proportionality:

\begin{equation}
\delta_{TDV} \propto M_S \cdot a_S^{-1/2}
\end{equation}

As a final note, we point out that in \S3.3 we discussed how increasing $e_S$ tends to decrease the TTV amplitude.  For the TDV signal, the opposite is true, increasing $e_S$ tends to increase the TDV signal.

\subsection{TTV \& TDV as complementary methods}

From equations (13) and (18), it is clear that the ratio of TDV to TTV should be able to eliminate $M_S$ and we can directly measure the orbital seperation of the exomoon $a_S$ and hence $M_S$.  The introduction of TDV allows for the precise measurement of $M_S$ without any assumption on the orbital separation.  Using Kepler's Third Law, it is then possible to derive the exomoon's orbital period, which was shown in \S3.1 to be unattainable from TTV alone.  In appendix D, it is shown that if we assume $e_S \simeq 0$ (but $e_P$ can take any value between zero and unity), then $\eta = \delta_{TDV}/\delta_{TTV}$ is approximately given by:

\begin{equation}
\eta = \frac{\delta_{TDV}}{\delta_{TTV}} \simeq  \frac{2 \pi \bar{\tau}}{P_P} \cdot \frac{\sqrt{3}}{\chi^{3/2}}
\end{equation}

Therefore one may determine $\chi$ and hence $a_S$ (and $M_S$) if one assumes the exomoon has zero eccentricity.  This assumption is supported based on a study by \citet{dom06} and the pattern observed amongst large ($>500$km) moons in our own Solar System.

Another major advantage of TDV is that the signal should lag TTV by a $\pi/2$ phase difference, originating from the fact TTV is a spatial effect whereas TDV is a velocity effect.  Unfortunately, the direction and value of the phase shift remains unchanged between prograde and retrograde satellites.

Combining TTV and TDV should allow for a much more significant confirmation of a potential exomoon than just using TTV alone.  The phase difference can be seen in the example waveforms in figure 3, where we plot the TTV and TDV waveforms due to a hypothetical exomoon of $1 M_{\bigoplus}$ around a GJ436b for $e_S = 0$, 0.3 and 0.6.

\begin{figure}
\begin{center}
\includegraphics[width=8.4 cm]{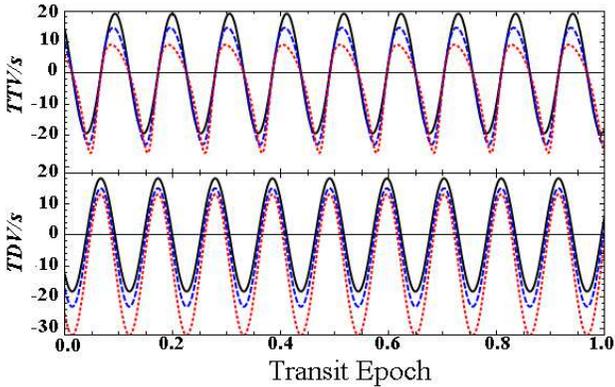}
\caption{\emph{The TTV and TDV waveforms due to a hypothetical $1 M_{\bigoplus}$ exomoon around GJ436b.  We show a) the effects of exomoon eccentricity: the solid line gives $e_S = 0$, the dashed line $e_S = 0.3$ and the dotted line $e_S = 0.6$ b) TTV leads TDV by a $\pi/2$ phase difference c) the high frequency nature of both waves (as discussed in \S3.1) is apparent d) increasing $e_S$ tends to decrease the TTV amplitude and increase the TDV amplitude (as discussed in \S3.3 \& \S 4.2)}} \label{fig:fig3}
\end{center}
\end{figure}

In Table 1, we predict the TTV and TDV rms amplitudes for a variety of known transiting planets.  We include the effects of the known planetary eccentricity, and consider a $1 M_{\bigoplus}$ exomoon with $\chi = 1/3$ and $e_S = 0$.

\begin{table}
\caption{\emph{Predicted TTV \& TDV rms amplitudes due to a $1 M_{\bigoplus}$ exomoon, for a selection of the best candidate transiting planets.  System parameters are taken from various references, which are shown.}} 
\centering 
\begin{tabular}{c c c c} 
\hline\hline 
Planet & $\delta_{TTV}$/s & $\delta_{TDV}$/s & Reference \\ [0.5ex] 
\hline 
GJ436b & 14.12 & 13.68 & \citet{alo08} \\
CoRoT-Exo-4b & 7.58 & 9.15 & \citet{aig08} \\
OGLE-TR-111b & 4.63 & 7.32 & \citet{dia08} \\
HAT-P-1b & 4.58 & 6.82 & \citet{joh08} \\
HD149026b & 3.61 & 9.76 & \citet{win07} \\
Lupus-TR-3b & 3.28 & 5.19 & \citet{wel08} \\
WASP-7b & 3.26 & 5.88 & \citet{hel08} \\
TrES-1b & 3.04 & 5.95 & \citet{wina07} \\
HD17156b & 3.07 & 1.06 & \citet{bar07} \\
HD209458b & 2.97 & 5.95 & \citet{kip08} \\
XO-5b & 2.65 & 4.69 & \citet{bur08} \\
HAT-P-4b & 2.54 & 8.34 & \citet{kov07} \\
HD189733b & 1.52 & 2.96 & \citet{winb07} \\ 
XO-3b & 0.41 & 0.87 & \citet{win08} \\ [1ex]
\hline\hline 
\end{tabular}
\label{table:nonlin} 
\end{table}

In most cases, the TDV amplitude is actually greater than the TTV amplitude.  We do however point out that the error in the transit duration is typically 2-3 times that of the mid-transit time (e.g. \citet{alo08}).  Nevertheless, the signal should be detectable, especially with future telescopes like Kepler\footnote{see http://kepler.nasa.gov/sci/}.  It is interesting to note that timing errors of just under 1 second could detect a $1 M_{\bigoplus}$ around a $11.79 M_{J}$ exoplanet in the case of XO-3b, a mass ratio of $\sim 3 \times 10^{-4}$.

\subsection{A hypothetical habitable exomoon}

Consider a hypothetical exoplanet which is identical to GJ436b except it has a period of $P_P = 35.7$ days and is on a circular orbit, putting it into the habitable zone of GJ436 with $T_{eq} \simeq 300$K.  Although the Neptune-like planet itself would not be an ideal place to search for life, an Earth mass exomoon would be.  Suppose there is a 1$M_{\bigoplus}$ exomoon with this planet on a circular orbit and $\chi=0.25$ and hence a period of $P_S \simeq 2.5$ days.  Could we detect such a provocative target using timing effects alone?

The predicted rms TTV amplitude would be 138s and the TDV amplitude would be 60s.  For GJ436b, \citet{alo08} used the 1.52m Telescopio Carlos S\'{a}nchez telescope and achieved a timing accuracy of $\sim 13$ seconds and a duration error of $\sim 50$ seconds.  This suggests that the detection of the exomoon should be presently possible through TTV from the ground, and feasible with TDV in the near future.  This illustrates that even ground-based instruments could detect an Earth-like body in the habitable zone using timing effects.

\section{Discussion \& Conclusions}

We have presented an updated model for the TTV signal due to an exomoon to include the effects of orbital eccentricity in both the exoplanet and the exomoon.  From the updated TTV model, we draw the following conclusions:

\begin{enumerate}
\item[(a)] TTV is degenerated in that it can only determine $M_S \times a_S$, where $M_S$ is the exomoon mass and $a_S$ is the exomoon's orbital radius.
\item[(b)] The TTV due to an exomoon can be significantly enhanced for exoplanets of $e_P > 0$ and $\varpi_P \sim 270^{\circ}$. However, it remains unclear how dynamically stable such exomoons would be.
\item[(c)] The TTV frequency will be always be greater than the sampling frequency of once every transit, implying we can only determine a set of possible harmonic frequencies for the exomoon's period, $P_S$.
\end{enumerate}

An exomoon is predicted to have another detectable timing effect on a transit in the form of transit duration variation (TDV).  We have derived an equation for predicting the TDV amplitude and drawn the following conclusions:

\begin{enumerate}
\item[($\tilde{\textrm{a}}$)] The ratio of TDV to TTV allows for the separate determination of $M_S$ and $a_S$; thus solving the `inverse problem' for exomoons.
\item[($\tilde{\textrm{b}}$)] The TDV signal is of a similar order of magnitude to the TTV signal.
\item[($\tilde{\textrm{c}}$)] the TDV signal is also enhanced for $e_P >0$ and $\varpi_P \sim 270^{\circ}$.
\item[($\tilde{\textrm{d}}$)] TDV lags TTV by a $90^{\circ}$ phase difference, making it an excellent complementary technique for exomoon detection.
\end{enumerate}

We also find that current ground-based telescopes could detect a $1 M_{\bigoplus}$ exomoon in the habitable zone around a Neptune-like exoplanet.  The author would therefore encourage observers to produce not only their mid-transit times, but also transit durations for each transit, rather than composite lightcurve durations.  This will allow constraints to be placed on the presence of exomoons around such planets.

\section*{Acknowledgments}

David M. Kipping is supported by STFC and UCL.  The author would like to thank Giovanna Tinetti and Alan Aylward for their support throughout this research.  The author would also like to thank Ignasi Ribas, Steve Fossey and Jean-Philippe Beaulieu for their input during technical discussions on this project.

\appendix

\section{TTV rms amplitude}

The root-mean-square amplitude of a waveform with displacement $\textrm{TTV}(f_W)$ as a function of some variable $f_W$ may be written as:

\begin{equation}
\delta_{TTV} = \sqrt{\frac{1}{2\pi} \int_0^{2 \pi} [\textrm{TTV}(f_W)]^2 \textrm{ d}f_W}
\end{equation}

where $f_W$ is true anomaly of the planet during its orbit around the planet-moon barycentre and is the only variable we integrate over and thus we are assuming:
\begin{itemize}
\item $e_P$ \& $\varpi_P$ do not change over the measurement time-scale of the TTV effect.
\item There are no other secular changes in the planet's orbit.
\item $e_S$  does not change over the measurement time-scale of the TTV effect.
\item $\varpi_S$ takes the same value when sampled once every planetary orbital period, but $f_S$ and $f_W$ do not.
\end{itemize}

In our case, the function $TTV(f_W)$ is given by the projected distance between the planet and the planet-moon barycentre divided by the projected velocity of the barycentre, as a function of the planet's true anomaly around the planet-moon barycentre, $f_W$.

\begin{equation}
\textrm{TTV}(f_W) = \frac{x_2'(f_W,e_S,\varpi_S)}{v_{B\bot}(e_P,\varpi_P)}
\end{equation}

Where $x'_2(f_W,e_S,\varpi_S)$ is the projected displacement of the planet away from the planet-moon barycentre (see figure 1), as a function of $f_W$.

To find $x'_2(f_W,e_S,\varpi_S)$, we take a similar approach to that of \citet{kip08}, where the author considered an initial frame $S'$ and then made a series of transformations.  If we start out with the same setup as the cited author, we have an ellipse located with the centre at the origin of an $x'$-$y'$ plot and the barycentre located at $(a_W e_W,0)$.  We account for the position of periastron (defined in figure 1) by rotating the ellipse counter-clockwise by an angle $\varpi_W$ about the $z'$-axis.  This gives us the $S_1'$ frame:

\begin{align}
x_1' &= x' \cos \varpi_W - y' \sin \varpi_W \\
y_1' &= x' \sin \varpi_W + y' \cos \varpi_W
\end{align}

To match the figure, we require a translation to place the planet-moon barycentre at the origin.  After applying this translation, we have found the desired $S_2'$ frame.

\begin{align}
x_2' &= x_1' - a_W e_W \cos \varpi_W \\
y_2' &= y_1' - a_W e_W \sin \varpi_W
\end{align}

Note that $e_W$, the eccentricity of the planet's orbit around the planet-moon barycentre, must be equal to $e_S$.  However, $\varpi_W$ will have a phase difference of $\pi$ relative to $\varpi_S$.  In the defined ellipse, the distance between the barycentre and the centre of the planet, $r_W$ as a function of true anomaly, $f_W$, is given by:

\begin{equation}
r_W(f_W) = \frac{a_W (1-e_W^2)}{1+e_W \cos f_W}
\end{equation}

With this equation, we may define the $x'$ and $y'$ position of the centre of the planet as a function of $f_W$.

\begin{align}
x' &= a_W e_W + r_W(f_W) \cdot \cos f_W \\
y' &=  r_W(f_W) \cdot \sin f_W
\end{align}

Having defined $x'(f_W)$ and $y'(f_W)$, we have also acquired $x_2'(f_W)$, using equation (A5).  Putting $x_2'(f_W)$ into equation (A2) and integrating between the limits 0 and $2 \pi$, as dictated by (A1), we find:

\begin{equation}
\delta_{TTV} = \frac{a_W \cdot \zeta_{T}(e_S,\varpi_S)}{\sqrt{2} \cdot v_{B\bot}(e_P,\varpi_P)}
\end{equation}

where we define $\zeta_T(e_S,\varpi_S)$ by:

\begin{equation}
\zeta_{T} = \frac{(1-e_S^2)^{1/4}}{e_S} \sqrt{e_S^2 + \cos (2 \varpi_S) (2 (1-e_S^2)^{3/2} - 2 + 3 e_S^2)}
\end{equation}

Note, that we have replaced $e_W$ with $e_S$, since they are equivalent, and $\varpi_W$ with $\varpi_S$ since the $\pi$ phase difference between them does not affect this expression.

$a_W$ may be re-written as:

\begin{equation}
a_W = a_S \frac{M_S}{M_{PRV}}
\end{equation}

And $a_S$ may be expressed in terms of the Hill radius, $d_{max}$.

\begin{equation}
a_S = \chi d_{max} = \chi a_P \Big(\frac{M_{PRV} -M_S}{3 M_*}\Big)^{1/3}
\end{equation}

In this expression, $a_P$ may also be re-written using Kepler's Third Law:

\begin{equation}
a_P = \Big(\frac{G (M_{PRV}+M_*) P_P^2}{4 \pi^2}\Big)^{1/3}
\end{equation}

Finally giving:

\begin{equation}
a_W = \chi \cdot \frac{M_S}{M_{PRV}} \cdot \Big(\frac{G (M_{PRV}+M_*) (M_{PRV}-M_S) P_P^2}{12 \pi^2 M_*}\Big)^{1/3}
\end{equation}

Let us now turn our attention to the expression for $v_{B\bot}(e_P,\varpi_P)$.  If we assume the time it takes for the planet to cross a distance $a_W$ is small compared to the period of the orbit, then we may assume $v_{B\bot}$ does not vary.  However, it is only the perpendicular component of $v_B$, which we label as $v_{B\bot}$, which dictates the TTV effect.  In the case of a circular orbit, it is easy to see that $v_{B\bot} = v_B$, but for eccentric obits, this is not the case.

Consider the orbit of the planet-moon barycentre around the host star.  To find the perpendicular component of the barycentre's orbital velocity, we must find the gradient of the tangent to an ellipse which has been rotated for the position of periastron, $\varpi_P$, and then translated so that the star is at the origin.

We now consider an $\tilde{x}$-$\tilde{y}$ plot in a frame $\tilde{S}$ with an ellipse centered at the origin and the star located at coordinates $(a_P e_P, 0)$.  We then apply a counter-clockwise rotation about the $\tilde{z}$-axis by an angle $\varpi_P$ into the $\tilde{S_1}$ frame.  We then apply the translation to reach the $\tilde{S_2}$ frame.  The equations are identical to (A3) \& (A4) and (A5) \& (A6) except we replace the $W$ subscript with $P$.  These equations may be re-written to make $\tilde{x}$ \& $\tilde{y}$ the subject as follows:

\begin{align}
\tilde{x} & = -\tilde{x_2} \cos  \varpi_P - \tilde{y_2} \sin  \varpi_P - a_P e_P \\
\tilde{y} &= \tilde{x_2} \sin \varpi_P - \tilde{y_2} \cos \varpi_P
\end{align}

Note that we have assumed the orbital eccentricity of the planet-moon barycentre around the host star is equivalent to the orbital eccentricity of the planet around the host star.  Equations (A16) \& (A17) must satisfy the standard equation of an ellipse:

\begin{equation}
\frac{\tilde{x}^2}{a_P^2} + \frac{\tilde{y}^2}{a_P^2 (1-e_P^2)} = 1
\end{equation}

We substitute equations (A16) \& (A17) into equation (A18) and then differentiate implicitly with respect to $\tilde{x_2}$.  We then rearrange to make d$\tilde{y_2}$/d$\tilde{x_2}$ the subject.  Finally, we substitute for $\tilde{x_2}$ and $\tilde{y_2}$ using equations (A5) \& (A6) to get:

\begin{equation}
\frac{\textrm{d}\tilde{y_2}}{\textrm{d}\tilde{x_2}} = \frac{\tilde{y} \sin \varpi_P - (1-e_P^2) \tilde{x} \cos  \varpi_P}{\tilde{y} \cos \varpi_P + (1-e_P^2) \tilde{x} \sin \varpi_P}
\end{equation}

$\tilde{x}(f_P)$ and $\tilde{y}(f_P)$ are known through the equations (A7), (A8) and (A9) except the subscript $W$ should be replaced by $P$ in these three expressions.

The angle the gradient of the tangent makes to the $+\tilde{x}$ axis is given by:

\begin{equation}
\tilde{\theta}(f_P) = \arctan \Big[\frac{\textrm{d}\tilde{y_2}}{\textrm{d}\tilde{x_2}}\Big]
\end{equation}

And finally we may express the perpendicular component of the barycentric velocity as:

\begin{equation}
v_{B\bot} = v_B \cos [\tilde{\theta}(f_P)]
\end{equation}

$f_P$ is the true anomaly of the planet-moon barycentre during transit and is given by $f_P = \pi/2 - \varpi_P$.  Performing the necessary substitutions we find:

\begin{equation}
\cos \tilde{\theta}(f_P)= \cos \Big[\arctan \Big(\frac{-e_P \cos \varpi_P}{1+e_P \sin \varpi_P}\Big)\Big]
\end{equation}

If we take the limit of the expression (A24) in the case of $e_P \rightarrow 0$, we get the expected result of simply $v_B$.  $v_B$ is also a function of $e_P$ and $\varpi_P$ and there is an obvious desideratum to create a single parameter which absorbs all the effects of planetary eccentricity.  Thus we write $v_B$ as:

\begin{equation}
v_B = \sqrt{\frac{G (M_{PRV} + M_*)}{a_P}} \cdot \sqrt{\frac{2 (1+e_P \sin \varpi_P)}{(1-e_P^2)}-1}
\end{equation}

And we may now create a single parameter to absorb the effects of planetary eccentricity, $\Upsilon$:

\begin{equation}
\Upsilon = \cos \Big[\arctan \Big(\frac{-e_P \cos \varpi_P}{1+e_P \sin \varpi_P}\Big)\Big] \cdot \sqrt{\frac{2 (1+e_P \sin \varpi_P)}{(1-e_P^2)}-1}
\end{equation}

Thus:
\begin{equation}
v_{B\bot} = \Upsilon(e_P,\varpi_P) \cdot \sqrt{\frac{G (M_{PRV} + M_*)}{a_P}}
\end{equation}

The final expression may be written as:

\begin{equation}
\delta_{TTV} = \frac{a_W \sqrt{a_P}}{\sqrt{2} \sqrt{G (M_{PRV} + M_*)}} \cdot \frac{\zeta_{T}(e_S,\varpi_S)}{\Upsilon(e_P,\varpi_P)}
\end{equation}

where $a_W$ is given by equation (A15), $a_P$ is given by equation (A14), $\zeta_T(e_S,\varpi_S)$ is given by equation (A11), $\Upsilon(e_P,\varpi_P)$ is given by equation (A27).  An alternative expression of (A26) is given by:

\begin{equation}
\delta_{TTV} = \frac{1}{\sqrt{2}} \cdot \frac{a_P^{1/2} a_S M_S M_{PRV}^{-1}}{\sqrt{G (M_* + M_{PRV})}} \cdot \frac{\zeta_{T}(e_S,\varpi_S)}{\Upsilon(e_P,\varpi_P)}
\end{equation}

It is therefore clear that our derived equation is equivalent to that of previous authors in the case of circular orbits.  It is also clear that $\delta_{TTV} \propto M_S a_S$.  

For completion and scaling purposes, we also derive the TTV amplitude in terms of just masses and periods by using Kepler's Third Laws.

\begin{equation}
\delta_{TTV} = \frac{P_P}{2 \pi} \cdot \frac{\chi}{3^{1/3} \sqrt{2}} \cdot Z_T(M_*,M_{PRV},M_S) \cdot \frac{\zeta_{T}(e_S,\varpi_S)}{\Upsilon(e_P,\varpi_P)}
\end{equation}

where $Z_T(M_*,M_{PRV},M_S)$ is the TTV mass function given by:

\begin{equation}
Z_T = \Big(\frac{M_S^6 (M_{PRV}-M_S)^2}{M_{PRV}^6 M_*^2}\Big)^{1/6} \simeq \Big(\frac{M_S^6}{M_*^2 M_{PRV}^4}\Big)^{1/6}
\end{equation}

Where the approximation is based on $M_{PRV} \gg M_S$.

\section{The $P_S/P_P$ Ratio}

From Kepler's Third Law, we may write:

\begin{equation}
P_S = \Big(\frac{4 \pi^2 a_S^3}{G M_{PRV}}\Big)^{1/2}
\end{equation}

where $a_S$ is some fraction, $\chi$, of the Hill radius:
\begin{equation}
a_S = \chi \cdot a_P \cdot \Big(\frac{M_{PRV}-M_S}{3 M_*}\Big)^{1/3} \simeq \chi \cdot a_P \cdot \Big(\frac{M_{PRV}}{3 M_*}\Big)^{1/3}
\end{equation}

Putting (B1) and (B2) together:
\begin{equation}
P_S = \Big(\frac{4 \pi^2 \chi^3 a_P^3}{3 G M_*}\Big)^{1/2}
\end{equation}

In comparison, the orbital period of the planet around the host star is given by:
\begin{equation}
P_P = \Big(\frac{4 \pi^2 a_P^3}{G (M_*+M_{PRV})}\Big)^{1/2} \simeq \Big(\frac{4 \pi^2 a_P^3}{G M_*}\Big)^{1/2}
\end{equation}

Therefore we may write:
\begin{equation}
\frac{P_S}{P_P} \simeq \sqrt{\frac{\chi^3}{3}}
\end{equation}

\section{Derivation of the TDV Amplitude}
\subsection{Derivation of the circular form}

We will follow the same definition for TDV as for TTV, where TDV is equal to the observed - calculated (O-C) transit duration.  The calculated (or expected) transit duration, $\tau$, is given by:

\begin{equation}
\bar{\tau} = \frac{X}{v_{B\bot}}
\end{equation}
Where $X$ is the distance the planet has to cross in order to complete the transit and $v_B$ is the velocity of the planet during transit, as given by Kepler's Third Law.

In the case of an additional moon orbiting the planet, the velocity of the planet now has two components, the velocity of the planet-moon barycentre ($v_B$) and the \emph{wobble} velocity due to the perturbation of the moon.  We remind the reader that we assume coplanar orbits with $i=90^{\circ}$.

\begin{equation}
\tau(f_W) = \frac{X}{v_{B\bot} + v_{W\bot}(f_W)}
\end{equation}

Therefore, TDV is defined as:

\begin{equation}
\textrm{TDV}(f_W) = \bar{\tau} - \tau(f_W) = \Big(\frac{v_{B\bot}}{v_{B\bot}+v_{W\bot}(f_W)} - 1\Big) \cdot \bar{\tau}
\end{equation}

If $v_{B\bot} \gg v_{W\bot}$, then we may write:

\begin{equation}
\textrm{TDV}(f_W) \simeq - \frac{v_{W\bot}(f_W)}{v_{B\bot}} \cdot \bar{\tau}
\end{equation}

If we assume a circular orbit, then the velocity of the planet around the barycentre of the planet-moon system, the \emph{wobble} velocity, is given by:

\begin{equation}
v_{W} = \frac{2 \pi a_W}{P_S} = \frac{2 \pi M_S a_S}{M_{PRV}} \cdot \frac{1}{P_S}
\end{equation}

Using equation (B1), we may write:

\begin{equation}
v_{W} =  \sqrt{\frac{G M_S^2}{a_S M_{PRV}}}
\end{equation}

However, we want the perpendicular component of this velocity, $v_{W\bot}$ as a function of true anomaly, $f_W$.  For a circular orbit, it is trivial to show that the variation will be sinusoidal and hence the rms amplitude of $v_{W\bot}$ is given by:

\begin{equation}
|v_{W\bot}| = \frac{1}{\sqrt{2}}\Big(\frac{G M_S^2}{a_S M_{PRV}}\Big)^{1/2}
\end{equation}

Let us now consider the velocity of the planet-moon barycentre around the host star.  For a circular orbit, the velocity of the barycentre is given by:

\begin{equation}
v_{B} = v_{B\bot} = \frac{2 \pi a_P}{P_P} \simeq \Big(\frac{2 \pi G M_*}{P_P}\Big)^{1/3}
\end{equation}

Once again, due to the circular orbit, the perpendicular component of the velocity is the same as the absolute value during transit.  The ratio of $v_{W\bot}/v_{B\bot}$ multiplied by the transit duration will be the TDV signal.  

\begin{equation}
\textrm{TDV}(f_W) = -\frac{v_{W\bot}(f_W)}{v_{B\bot}} \cdot \bar{\tau} = -R \cdot \bar{\tau}
\end{equation}

In the case of circular orbits, $R$ is therefore given by:

\begin{equation}
R = \frac{1}{\sqrt{2}}\Big(\frac{G M_S^2}{a_S M_{PRV}}\Big)^{1/2} \Big(\frac{2 \pi G M_*}{P_P}\Big)^{-1/3}
\end{equation}

From equation (C5), we may instantly infer that the TDV effect is $\propto M_S a_S^{-1/2}$, in contrast to the TTV effect which is $\propto M_S a_S$.  In the case of $M_* \gg M_{PRV}$ and $M_{PRV} \gg M_S$, $a_S$ is given by:

\begin{equation}
a_S \simeq \chi \cdot \Big(\frac{G M_{PRV} P_P^2}{12 \pi^2}\Big)^{1/3}
\end{equation}

And feeding this into the equation for $R$ we get:

\begin{equation}
R = \frac{1}{\sqrt{2}} \cdot \frac{1}{\sqrt{\chi}} \cdot \Big(\frac{3 M_S^6}{M_*^2 M_{PRV}^4}\Big)^{1/6}
\end{equation}

Therefore the TDV amplitude for a planet, on a circular orbit, with a single moon, which is also on a circular orbit, is given by:

\begin{equation}
\delta_{TDV} \simeq \frac{\tau}{\sqrt{\chi}} \cdot \frac{3^{1/6}}{\sqrt{2}} \cdot Z_D(M_*,M_{PRV},M_S)
\end{equation}

where the TDV mass function is defined by:

\begin{equation}
 Z_D(M_*,M_{PRV},M_S) \simeq \Big(\frac{M_S^6}{M_*^2 M_{PRV}^4}\Big)^{1/6}
 \end{equation}

\subsection{Derivation of the eccentric form}

Here, we derive the TDV rms amplitude in the case of eccentric orbits.  We use the definition of TDV given by equation (C4).  The orbital velocity of the moon around the barycentre of the planet-moon system for a circular orbit is given by:

\begin{equation}
v_{W} = \sqrt{\frac{G M_S^2}{a_S M_{PRV}}} = \sqrt{\frac{G M_S^3}{a_W M_{PRV}^2}} = \sqrt{\frac{\mu_W}{a_W}}
\end{equation}

In the case of an eccentric orbit, this velocity becomes a function of true anomaly, $f_W$.

\begin{equation}
v_{W}(f_W) = \mu_W^{1/2} \cdot \Big(\frac{2}{r_W(f_W)}-\frac{1}{a_W}\Big)^{1/2}
\end{equation}

The TDV signal has a waveform governed by the ratio of the perpendicular component of the planet's wobble velocity to the perpendicular component of the velocity of the planet-moon barycentre around the host star, multiplied by the duration of the transit.  

\begin{equation}
\textrm{TDV}(f_W) = \frac{\bar{\tau}}{v_{B\bot}} \cdot v_{W,\bot}(f_W)
\end{equation}

In order to find the rms amplitude of this signal, we need to derive $v_{W,\bot}(f_W)$, which is not the same as $v_W(f_W)$.  To do this, we need to find the gradient of the tangent of the planet's position along our transformed ellipse from figure A.  To find the gradient of the tangent to the ellipse at any true anomaly, $f_W$, we need to implicitly differentiate the equation for the ellipse in the $S_2'$ frame.  Equations (A5) and (A6) may be rewritten making $x'$ and $y'$ the subject:

\begin{align}
x' &= -x_2' \cos  \varpi_W - y_2' \sin  \varpi_W - a_W e_W \\
y' &= x_2' \sin \varpi_W - y_2' \cos  \varpi_W
\end{align}

These two equations must satisfy the standard equation of the ellipse:

\begin{equation}
\frac{x'^2}{a_W^2}+\frac{y'^2}{a_W^2(1-e_W^2)} = 1
\end{equation}

We substitute (C18) and (C19) into (C20) and then differentiate implicitly with respect to $x_2'$.  We arrange the equation to make d$y_2'$/d$x_2'$ the subject.  Finally, we replace the $x_2'$ and $y_2'$ terms using equations (A5) and (A6) and find:

\begin{equation}
\frac{\textrm{d}y_2'}{\textrm{d}x_2'} = \frac{y' \sin  \varpi_W - (1-e_W^2) x' \cos \varpi_W}{y' \cos \varpi_W + (1-e_W^2) x' \sin  \varpi_W}
\end{equation}

$x'(f_W)$ and $y'(f_W)$ are known and so we have found the gradient of our tangent.  The angle of the tangent is given by:

\begin{equation}
\theta'(f_W) = \arctan \Big[\frac{\textrm{d}y_2'}{\textrm{d}x_2'}\Big]
\end{equation}

And finally we may express $v_{W,\bot}(f_W)$ as:

\begin{equation}
v_{W,\bot}(f_W) = v_W(f_W) \cdot \cos [\theta'(f_W)]
\end{equation}

We now take the rms of this waveform in the way detailed in appendix A and find:

\begin{equation}
\delta_{TDV} = \frac{\tau}{v_{B\bot}} \cdot \sqrt{\frac{\mu_W}{2 a_W}} \cdot \zeta_D(e_S,\varpi_S)
\end{equation}

where

\begin{equation}
\zeta_D(e_S,\varpi_S) = \sqrt{\frac{1 + e_S^2 - e_S^2 \cos (2 \varpi_S)}{1-e_S^2}}
\end{equation}

Note that we have replaced $e_W$ by $e_S$ since they are equivalent and similarly for $\cos 2\varpi_W$ and $\cos 2\varpi_S$.  Taking the limit of this expression as $e_S \rightarrow 0$ gives the expected result of $\tau \sqrt{\mu_W}/v_B \sqrt{2 a_W}$.  $v_{B\bot}$ has already been derived in appendix A and equation (A28).

\begin{equation}
v_{B\bot} = \Upsilon(e_P,\varpi_P) \cdot \sqrt{\frac{G (M_{PRV} + M_*)}{a_P}}
\end{equation}

Using this and equation (C9) we may re-write $\delta_{TDV}$ as:

\begin{equation}
\delta_{TDV} = \sqrt{\frac{a_P}{a_S}} \cdot \sqrt{\frac{M_S^2}{M_{PRV} (M_{PRV} + M_*)}} \cdot \frac{\bar{\tau}}{\sqrt{2}} \cdot \frac{\zeta_D(e_S,\varpi_S)}{\Upsilon(e_P,\varpi_P)} 
\end{equation}

This once again demonstrates the effect's proportionality of $\propto M_S a_S^{-1/2}$.  Using equation (A13), the fraction $a_P/a_S$ may be substituted for and we find:

\begin{equation}
\delta_{TDV} = \frac{\tau}{\sqrt{\chi}} \cdot \frac{3^{1/6}}{\sqrt{2}} \cdot Z_D(M_*,M_{PRV},M_S) \cdot \frac{\zeta_D(e_S,\varpi_S)}{\Upsilon(e_P,\varpi_P)} 
\end{equation}

where
\begin{equation}
Z_D(M_*,M_{PRV},M_S) = \Big(\frac{M_* M_S^6}{M_{PRV}^3 (M_{PRV} + M_*)^3 (M_{PRV} - M_S)}\Big)^{1/6}
\end{equation}

This general expression can be shown to be equivalent to our approximate form with the same approximations made.

\section{Ratio of TDV to TTV amplitude}

In this section we derive the ratio of the TDV amplitude to the TTV amplitude.  This will allow us to analytically quickly see which signal is stronger for any new system we come across and solve for the exomoon mass exactly.  We have:

\begin{equation}
\delta_{TTV} = \frac{P_P}{2 \pi} \cdot \frac{\chi}{3^{1/3} \sqrt{2}} \cdot Z_T(M_*,M_{PRV},M_S) \cdot \frac{\zeta_{T}(e_S,\varpi_S)}{\Upsilon(e_P,\varpi_P)}
\end{equation}

\begin{equation}
\delta_{TDV} = \frac{\bar{\tau}}{\sqrt{\chi}} \cdot \frac{3^{1/6}}{\sqrt{2}} \cdot Z_D(M_*,M_{PRV},M_S) \cdot \frac{\zeta_D(e_S,\varpi_S)}{\Upsilon(e_P,\varpi_P)} 
\end{equation}

We make the assumptions that $M_* \gg M_{PRV}$ and $M_{PRV} \gg M_S$.  In this case, the TTV mass function, $Z_T$, and the TDV mass function, $Z_D$, are equal.  We also make the assumption that moon is on a circular orbit and so $\zeta_T = \zeta_D = 1$.  Furthermore, the planet's eccentricity parameter, $\Upsilon$ will cancel out and so we need not make the assumption $e_P = 0$.  This gives us the ratio of TDV to TTV, $\eta$ to be:

\begin{equation}
\eta = \frac{\delta_{TDV}}{\delta_{TTV}} \simeq  \frac{2 \pi \bar{\tau}}{P_P} \cdot \frac{\sqrt{3}}{\chi^{3/2}} = \frac{2 \pi \bar{\tau}}{P_S}
\end{equation}

The masses have completely cancelled out and we are able to solve the equation for $P_S$ in a  completely rigourous way.

\bsp

\label{lastpage}


\begin{thebibliography}{99}
\bibitem[\protect\citeauthoryear{Agol et al.}{2005}]{ago05} Agol, E., Steffen, J., Sari, R. \& Clarkson, W., 2005, MNRAS, 359, 567
\bibitem[\protect\citeauthoryear{Aigrain et al.}{2008}]{aig08} Aigrain, S. et al., A\&A, 488, L43
\bibitem[\protect\citeauthoryear{Alonso et al.}{2008}]{alo08} Alonso, R., Barbieri, M., Rabus, M., Deeg, H. J., Belmonte, J. A., \& Almenara, J. M. 2008, A\&A, 363, 1081
\bibitem[\protect\citeauthoryear{Barbieri et al.}{2007}]{bar07} Barbieri, M. et al. A\&A, 476, L13
\bibitem[\protect\citeauthoryear{Barnes \& O'Brien}{2002}]{bar02} Barnes, J. W. \& O'Brien, D. P., 2002, ApJ, 575, 1087
\bibitem[\protect\citeauthoryear{Burke et al.}{2008}]{bur08} Burke, C. J. et al., ApJ, 686, 1331
\bibitem[\protect\citeauthoryear{Cabrera \& Schneider}{2007}]{cab07} Cabrera, J. \& Schneider, A\&A, 464, 1133
\bibitem[\protect\citeauthoryear{D\'{i}az et al.}{2008}]{dia08} D\'{i}az, R. F. et al. 2008, ApJ, 682, L49
\bibitem[\protect\citeauthoryear{Domingos et al.}{2006}]{dom06} Domingos, R. C., Winter, O. C. \& Yokoyama, T. 2006, MNRAS, 373, 1227
\bibitem[\protect\citeauthoryear{Ford \& Holman}{2007}]{for07} Ford, E. B. \& Holman, M. J., 2007, ApJ, 664, L51
\bibitem[\protect\citeauthoryear{Gillon et al.}{2007}]{gil07} Gillon, M., Triaud, A.H.M.J., Mayor, M., Queloz, D., Udry. S., \& North, P., 2008, A\&A, 485, 871
\bibitem[\protect\citeauthoryear{Hellier et al.}{2008}]{hel08} Hellier, C. et al. 2008, preprint (arXiv:0805:2600)
\bibitem[\protect\citeauthoryear{Holman \& Murray}{2005}]{hol05} Holman, M. J. \& Murray, N. W., Sci, 307, 5713
\bibitem[\protect\citeauthoryear{Johnson et al.}{2008}]{joh08} Johnson, J. A. et al. 2008, ApJ, 686, 649
\bibitem[\protect\citeauthoryear{Kane \& von Braun}{2008}]{kan08} Kane, S. R. \& von Braun, K. 2008, in press (arXiv:0808.1890)
\bibitem[\protect\citeauthoryear{Kipping}{2008}]{kip08} Kipping, D. M., 2008, MNRAS, 389, 1383
\bibitem[\protect\citeauthoryear{Kov\'{a}cs et al.}{2007}]{kov07} Kov\'{a}cs, G. et al. 2007, ApJ, 670, L41
\bibitem[\protect\citeauthoryear{P\'{a}l \& Kocsis}{2008}]{pal08} P\'{a}l, A. \& Kocsis, B. 2008, MNRAS, 389, 191
\bibitem[\protect\citeauthoryear{Sartoretti \& Schneider}{1999}]{sar99} Sartoretti, P. \& Schneider, J., 1999, A\&AS, 14, 550
\bibitem[\protect\citeauthoryear{Simon et al.}{2007}]{sim07} Simon, A., Szatm\'{a}ry, K., Szab\'{o}, Gy. M.,  2007, A\&A, 470, 727S
\bibitem[\protect\citeauthoryear{Szab\'{o}  et al.}{2006}]{sza06} Szab\'{o}, Gy. M., Szatm\'{a}ry, K., Div\'eki, Zs. \& Simon, A.,  2006, A\&A, 450, 395
\bibitem[\protect\citeauthoryear{Valencia  et al.}{2006}]{val06} Valencia, D., Sasselov, D. D. \& O'Connell, R. J.,  2006, Icarus, 181, 545
\bibitem[\protect\citeauthoryear{Weldrake et al.}{2008}]{wel08} Weldrake, D. T. F. et al. 2008, ApJ, 675, L37
\bibitem[\protect\citeauthoryear{Winn et al.}{2007}]{win07} Winn, J. N., Henry, G. W., Torres, G. \& Holman, M. J. 2007, ApJ, 675, 1531
\bibitem[\protect\citeauthoryear{Winn et al.}{2007a}]{wina07} Winn, J. N., Holman, M. J. \& Roussanova, A. 2007, ApJ, 657, 1098
\bibitem[\protect\citeauthoryear{Winn et al.}{2007b}]{winb07} Winn, J. N., Holman, M. J., Henry, G. W. et al. 2007, AJ, 134, 1707
\bibitem[\protect\citeauthoryear{Winn et al.}{2008}]{win08} Winn, J. N. et al. 2008, ApJ, 683, 1076
\end{thebibliography}
\end{document}